\documentclass[11pt,prd,aps,nofootinbib,floatfix]{revtex4}
\usepackage[utf8]{inputenc}
\usepackage{mathtools}
\usepackage{multirow}
\usepackage{amsmath}
\usepackage{amssymb,slashed}

\makeatletter

\providecommand{\tabularnewline}{\\}

\@ifundefined{textcolor}{}
{%
 \definecolor{BLACK}{gray}{0}
 \definecolor{WHITE}{gray}{1}
 \definecolor{RED}{rgb}{1,0,0}
 \definecolor{GREEN}{rgb}{0,1,0}
 \definecolor{BLUE}{rgb}{0,0,1}
 \definecolor{CYAN}{cmyk}{1,0,0,0}
 \definecolor{MAGENTA}{cmyk}{0,1,0,0}
 \definecolor{YELLOW}{cmyk}{0,0,1,0}
}

\usepackage{graphicx}\usepackage{epsfig}\usepackage{dsfont}\usepackage[usenames]{color}
\usepackage{ulem}
\usepackage{bigstrut}
\usepackage{slashed}
\usepackage{multirow}
\usepackage{subfigure}
\usepackage{color}
\usepackage{xcolor}

\allowdisplaybreaks

\makeatother

\begin{document}
\title{The strong coupling constants of doubly heavy baryons with heavy mesons}
\author{Xiao-Hui Hu$^{1}$~$^{2}$~\thanks{Corresponding author Email:~huxiaohui@cumt.edu.cn}, Quan-Yu Zhou$^{1}$~\thanks{Corresponding author Email:14224300@cumt.edu.cn}, Ye Xing$^{1}$~\thanks{Corresponding author Email:~xingye_guang@cumt.edu.cn}, Yu-Ji Shi$^{3}$~\thanks{Corresponding author Email:~shiyuji@ecust.edu.cn}}
\affiliation{$^{1}$ Department of Applied Physics, the School of Materials Science and physics,
China University of mining and technology, Xuzhou 221116, China}
\affiliation{$^{2}$ Lanzhou Center for Theoretical Physics and Key Laboratory
of Theoretical Physics of Gansu Province, Lanzhou University, Lanzhou
730000, China}
\affiliation{$^{3}$ School of Physics, East China University of Science and Technology, Shanghai 200237, China}
\begin{abstract}
In the experiment, only one doubly charmed baryon $\Xi_{cc}^{++}$ was successfully identified.
The LHCb collaboration is poised to discover other doubly heavy baryon, $\Xi_{cc}^{+}$ and $\Xi_{bc}^{+}$, necessitating a comprehensive study of doubly heavy baryons from various perspectives. Such investigations will elucidate the properties of the newly discovered states and facilitate the search for other doubly heavy baryons as predicted in quark model. In this paper, we explore the strong coupling between the doubly heavy baryon, singly heavy baryon, and heavy meson, ${\cal B}_{QQ^{\prime}}{\cal B}_{Q}M_{Q^{\prime}}$, using the light cone QCD sum rules approach. We analytically and numerically calculate the strong coupling, ${\cal B}_{QQ^{\prime}}{\cal B}_{Q}M_{Q^{\prime}}$, utilizing the light-cone distribution amplitudes of the singly heavy baryon ${\cal B}_{Q}$. The numerical predictions for these strong couplings are provided, along with the error analysis. Our theoretical findings can be employed to examine the strong and weak decay dynamics of doubly heavy baryons. Furthermore, our results may offer experimentalists a valuable tool for analyzing data on the strong coupling constants among hadronic multiplets.
\end{abstract}
\maketitle

\section{Introduction}
High-energy experiments such as LHCb, CMS, CDF, Belle and BaBar Collaborations have discovered many new hadrons with heavy quarks. These discoveries have spurred a lot of activities in both experimental and theoretical hadronic physics~\cite{Chen:2022asf}. The study of heavy baryons and mesons provides an important means to deepen our understanding of hadron structure, strong interaction dynamics and heavy-quark symmetry. 
The doubly heavy baryons with spin of 1/2 constitute a light flavor triplet ${\cal B}_{QQ^{\prime}}^{\mathbf{3}}$, wherein $Q^{(\prime)}$ represents either $c$ or $b$.
The search for these doubly heavy baryons has been an enduring challenge in the experiment.
Notably, the initial evidence of this state was unveiled in 2005 by the SELEX experiment. This discovery was characterized by the decay of $\Xi_{cc}^{+}$ into final states of $\Lambda_c^+ K^- \pi^+$ and $pD^+K^-$~\cite{SELEX:2002wqn,SELEX:2004lln}.
However, the existence of $\Xi^{+}_{cc}(3519)$, as reported by SELEX~\cite{SELEX:2002wqn,SELEX:2004lln}, has not been confirmed by subsequent experiments conducted by FOCUS~\cite{Ratti:2003ez}, BaBar~\cite{BaBar:2006bab}, LHCb~\cite{LHCb:2013hvt}, and Belle~\cite{Belle:2013htj}. Consequently, $\Xi^{+}_{cc}(3519)$ has not been listed in the Particle Data Group (PDG)~\cite{ParticleDataGroup:2024cfk}.
To date, only one candidate has been experimentally identified. The LHCb Collaboration detected a signal for the $\Xi^{++}_{cc}(3621)$ baryon in the $\Lambda^+_c K^-\pi^+\pi^+$ invariant mass~\cite{LHCb:2017iph}. The measured mass of the $\Xi_{cc}^{++}$ is $3620.6\pm1.5(\text{stat})\pm0.4(\text{syst})\pm0.3(\Xi_{c}^{+})\text{MeV}/c^{2}$, which aligns with theoretical predictions~\cite{Lewis:2001iz,Karliner:2014gca}.
This was confirmed by an independent decay mode $\Xi_{cc}^{++}\to\Xi_{c}^{+}\pi^{+}$ in 2018~\cite{LHCb:2018pcs}, and the lifetime of $\Xi^{++}_{cc} $ was measured to be $0.256\,^{+0.024}_{-0.022}(\text{stat.})\pm
0.014(\text{syst.})~\text{ps}$~\cite{Aaij:2018wzf}.
Precision measurement of the $\Xi_{cc}^{++}$ mass has also been performed by LHCb\cite{LHCb:2019epo}.
Certainly, future experimental efforts will continue to investigate the unexplained discrepancies in the parameters of these states, as well as further the search for additional doubly heavy baryons.
To date, the LHCb collaboration has conducted searches for various doubly heavy baryons, including $\Xi_{cc}^{+}$ ($ccd$)~\cite{LHCb:2019gqy,LHCb:2021eaf}, $\Omega_{cc}^+$ ($ccs$)~\cite{LHCb:2021rkb}, $\Xi_{bc}^{0}~(bcd)$~\cite{LHCb:2020iko}, $\Omega_{bc}^{0}~(bcs)$~\cite{LHCb:2021xba}, and $\Xi_{bc}^{+}~(bcu)$~\cite{LHCb:2022fbu}. However, only suggestive hints of signals were detected, implying that these states remain unobserved.
Future searches conducted at the LHCb, utilizing enhanced trigger conditions, an increased number of decay modes for doubly heavy baryons, and larger datasets, are anticipated to significantly improve the sensitivity of the doubly heavy baryons signal. 


Theoretical investigations on properties of the doubly heavy baryons, are necessary as their results can provide many inputs to experiments and help us better understand their internal structure and the dynamics of the QCD as the theory of the strong interaction. For the study of doubly heavy baryons, a number of comprehensive theoretical
researches have predicted their masses spectrum, lifetimes and other prospects
serving for the experimental detection.
The masses spectrum of the doubly heavy baryons have been calculated within the Lattice QCD~\cite{Mathur:2018epb,Mathur:2018rwu,Liu:2009jc}, the QCD sum rules (QCDSR) ~\cite{Bagan:1992za,Zhang:2008rt,Wang:2010hs,Aliev:2012ru,Aliev:2012iv,Azizi:2014jxa,Hu:2017dzi,Wang:2018lhz}, the chiral partner structure and heavy quark spin-flavor symmetry~\cite{Ma:2017nik}, covariant baryon chiral perturbation theory~\cite{Yao:2018ifh}, the unitarized coupled channel
approach based on chiral effective Lagrangians~\cite{Yan:2018zdt}, heavy quark-diquark symmetry~\cite{Brodsky:2011zs}.
The theoretical calculation of the lifetimes of doubly heavy baryons have been performed by using the operator product expansion (OPE)~\cite{Onishchenko:1999yu,Berezhnoy:2018bde} and the heavy quark expansion (HQE)~\cite{Cheng:2018mwu}.
The weak decays of doubly heavy baryons have been investigated within various nonperturbative approaches, including light cone sum rules (LCSRs) ~\cite{Shi:2019fph,Hu:2019bqj}, light front quark model (LFQM) ~\cite{Wang:2017mqp,Xing:2018lre,Zhao:2018mrg}, the flavor SU(3) symmetry~\cite{Wang:2017azm,Zhang:2018llc,Shi:2017dto}, the covariant confined quark model~\cite{Gutsche:2017hux}
the combination of factorization and pole model approaches~\cite{Sharma:2017txj,Dhir:2018twm}, the ab initio three-loop quark model~\cite{Gutsche:2018msz}, the combination of factorization hypothesis and the final-state interactions (FSIs)~\cite{Jiang:2018oak}.
The transition magnetic moments of doubly heavy baryon have been calculated in the heavy baryon chiral perturbation theory (HBChPT)\cite{Li:2017pxa,Liu:2018euh}, the light-cone sum rule~\cite{Ozdem:2018uue}.
The strong and radiative decays of the doubly heavy baryons have also studied \cite{Xiao:2017udy}.
The strong coupling constants of doubly heavy baryons with light meson have been obtained by means of light cone QCD sum rule~\cite{Olamaei:2020bvw,Azizi:2020zin,Rostami:2020euc}.

From the above, it is evident that the majority of research concentrates on the mass and weak decays of doubly heavy baryons. Notably, there exists a paucity of studies addressing their strong decays and the strong couplings between doubly heavy baryons and other hadrons. Furthermore, no known research explores the strong couplings of doubly heavy baryons with heavy mesons. Considering that the mass of doubly heavy baryons is below the mass threshold of single-flavor baryons and heavy mesons, they do not experience strong decay at the tree diagram level. Nevertheless, when evaluating the higher-order loop diagram calculations during the weak decay of doubly heavy baryons, techniques such as the final state interaction are used to investigate their non-leptonic weak decay. In these contexts, virtual heavy mesons become relevant, rendering the strong coupling between the doubly heavy baryon, singly heavy baryon, and heavy meson, denoted as ${\cal B}_{QQ^{\prime}}{\cal B}_{Q}M_{Q^{\prime}}$, a crucial input parameter.

Also the strong couplings provide an important tool to study the structure of baryons. So in this paper we will firstly investigate the strong couplings among the doubly heavy baryon, singly heavy baryon, and heavy meson, ${\cal B}_{QQ^{\prime}}{\cal B}_{Q}M_{Q^{\prime}}$, by using the light-cone sum rules approach with the SU(3) light flavor symmetry.  As we know, the light-cone QCD sum rules is extened from the QCD sum rules method. The QCDSR method is based on the fundamental QCD Lagrangian, and it takes into account the non-perturbative nature of the QCD vacuum. While in the LCSRs approach~\cite{Braun:1988qv,Chernyak:1990ag,Ball:1998je,Ball:2006wn,Ball:2004rg,Ball:1998kk,Ball:1998sk,Ball:1998ff,Ball:2007rt,Ball:2007zt,Wang:2007mc,Wang:2009hra,Aliev:2010uy,Sun:2010nv,Khodjamirian:2011jp,Han:2013zg,Offen:2013nma,Meissner:2013hya}, a light-cone variant is further introduced to conduct the OPE based on the twists of the operators near the light-cone $x^2 \sim 0$ instead of the short distance $x\sim 0$. 
All non-perturbative effects are automatically included in the matrix elements of non-local operators. The matrix elements are parameterized in terms of light-cone distribution amplitudes (LCDAs) rather than vacuum condensates~\cite{Balitsky:1989ry,Chernyak:1983ej,Colangelo:2000dp}. The non-perturbative parameters contained in the LCDAs are evaluated by using QCDSR, and therefore are universal.

This paper is organized as follows. In Sec.\ref{sec:framework}, we briefly introduce the calculation of the strong coupling ${\cal B}_{QQ^{\prime}}{\cal B}_{Q}M_{Q^{\prime}}$ within the LCSRs approach. In Sec.\ref{sec:numericalresults}, we give the numerical results of these strong couplings and analysis the dependence of the parameters of $M^2_{1,2}$ and $s_{th}$. In Sec.\ref{sec:conclusions}, we discuss the results and conclude this paper.

\section{Theoretical framework}
\label{sec:framework}
In the LCSR approach, the cornerstone is the correlation function. It can be calculated in two
different ways. In the timelike region, one can insert the complete set of hadronic states
with the same quantum numbers as the interpolating currents to extract and isolate
the ground states. It is called the phenomenological or physical side of the correlation function. In
the spacelike region, which is free of singularities, one can calculate the correlation function in terms of QCD level of freedom using OPE. It is known as the QCD or the theoretical side.
These two representations, which respectively are the real and imaginary parts of the
correlation function, can be matched via a Quark-Hadron duality to find the corresponding sum rule. The
divergences coming from the dispersion integral, as well as higher states and continuum,
are suppressed using the well-known method of Borel transformation and continuum
subtraction.

In this section, we will introduce the calculation of the strong couplings of doubly heavy baryon, singly heavy baryon anti-triplet and heavy meson ${\cal B}_{QQ^{\prime}}{\cal B}_{Q}M_{Q^{\prime}}$ within the framework LCSRs using the light cone distribution amplitudes of singly heavy baryon anti-triplet. In SubSec.~\ref{subsec:hadronlevel}, we derive the sum rules in hadron level.
In SubSec.~\ref{subsec:qcdlevel3}, we derive the sum rules in QCD level from the LCDAs of the singly heavy baryon anti-triplet ${\cal B}_{Q}$. 
In SubSec.~\ref{subsec:quark hadronquality}, we discussed the Quark -Hadron Duality.
\subsection{Hadron level}
\label{subsec:hadronlevel}
The two-point correlation function can be written as 
\begin{eqnarray}
  &  & \Pi_{\mu}(p,q)=i\int d^4xe^{iq\cdot x}\langle{\cal B}_{Q}(p)|T{j_{\mu}(x)\bar{j}_{{\cal B}_{QQ^{\prime}}}(0)}|0\rangle,\label{eq:correlation5}\\
  &  & \Pi_{5}(p,q)=i\int d^4xe^{iq\cdot x}\langle{\cal B}_{Q}(p)|T{j_{5}(x)\bar{j}_{{\cal B}_{QQ^{\prime}}}(0)}|0\rangle,\label{eq:correlationmu}
 \end{eqnarray}
where the pseudoscalar and vector currents can be defined as 
\begin{eqnarray}
j_{5}(x) & = & \bar{q}_{a}(x)i\gamma_{5}Q_{a}(x),\nonumber \\
j_{\mu}(x) & = & \bar{q}_{a}(x)\gamma_{\mu}Q_{a}(x).
\end{eqnarray}
And for $Q=Q^{\prime}=b,\ c$, the current of doubly heavy baryon can be given as 
\begin{eqnarray}
\bar{j}_{{\cal B}_{QQ^{\prime}}}(0) & = & \epsilon_{abc}\bar{q}_{c}\gamma_{\mu}\gamma_{5}(\bar{Q}_{b}(0)\gamma_{\mu}C\bar{Q}_{a}^{\prime T}(0)),
\end{eqnarray}
while for $Q=b,\ Q^{\prime}=c$, the current is
\begin{equation}
  \bar{j}_{{\cal B}_{bc}}(0)=\frac{1}{\sqrt{2}}\epsilon_{abc}\bar{q}_{c}\gamma_{\mu}\gamma_{5}(\bar{b}_{b}(0)\gamma_{\mu}C\bar{c}_{a}^{T}(0)+\bar{c}_{b}^{T}(0)\gamma_{\mu}C\bar{b}_{a}(0)).
\end{equation}
Here, $q=u,~d,~s$ represent the quark field, $a,~b,~c$ are color indexes. ${\cal B}_{QQ^{\prime}}$ and ${\cal B}_{Q}$ represent doubly heavy baryon and singly heavy baryon, respectively. Under the SU(3) flavor symmetry, their martixs can be written as following~\cite{Li:2012bt,Yan:1992gz,Hu:2020mxk}, 
\begin{eqnarray}
  &  & {\cal B}_{c}=\left(\begin{array}{ccc}
    0&\Lambda_{c}^{+}&\Xi_{c}^{+}\\
    -\Lambda_{c}^{+}&0&\Xi_{c}^{0}\\
    -\Xi_{c}^{+}&-\Xi_{c}^{0}&0
  \end{array}\right),~~~
  {\cal B}_{b}=\left(\begin{array}{ccc}
    0&\Lambda_{b}^{0}&\Xi_{b}^{0}\\
    -\Lambda_{b}^{0}&0&\Xi_{b}^{-}\\
    -\Xi_{b}^{0}&-\Xi_{b}^{-}&0
  \end{array}\right),\nonumber\\
  &  & {\cal B}_{cc}=\left(\Xi_{cc}^{++}~\Xi_{cc}^{+}~\Omega_{cc}^{+}\right), ~~{\cal B}_{bc}=\left(\Xi_{bc}^{+}~\Xi_{bc}^{0}~\Omega_{bc}^{0}\right),~~{\cal B}_{bb}=\left(\Xi_{bb}^{0}~\Xi_{bb}^{-}~\Omega_{bb}^{-}\right).\nonumber
 \end{eqnarray}
 In order to characterize our correlation function at the hadron level, we need to know the phenonmenological strong interaction effective lagrangian of the couplings~\cite{Li:2012bt,Yan:1992gz},  
\begin{eqnarray}
 &  & {\cal L}_{P{\cal B}_{Q}{\cal B}_{QQ^{\prime}}}=g_{P{\cal B}_{Q}{\cal B}_{QQ^{\prime}}}Tr[\bar{{\cal B}}_{Q}i\gamma_{5}P{\cal B}_{QQ^{\prime}}],\label{eq:LAPBB}\\
 &  & {\cal L}_{V{\cal B}_{Q}{\cal B}_{QQ^{\prime}}}=f_{1V{\cal B}_{Q}{\cal B}_{QQ^{\prime}}}Tr[\bar{{\cal B}}_{Q}\gamma_{\mu}V^{\mu}{\cal B}_{QQ^{\prime}}]+\frac{f_{2V{\cal B}_{Q}{\cal B}_{QQ^{\prime}}}}{m_{{{\cal B}_{Q}}}+m_{{{\cal B}_{QQ^{\prime}}}}}Tr[\bar{{\cal B}}_{Q}\sigma_{\mu\nu}\partial^{\mu}V^{\nu}{\cal B}_{QQ^{\prime}}],\label{eq:LAVBB}
\end{eqnarray}
where $g_{P{\cal B}_{Q}{\cal B}_{QQ^{\prime}}}$ and $f_{V{\cal B}_{Q}{\cal B}_{QQ^{\prime}}}$ both denote the strong coupling constants. 
And $P(V)$ denotes the pseudoscalar (vector) heavy meson, including $B^{(*)}$ and $D^{(*)}$ mesons, which can be listed as following,
\begin{eqnarray}
  &  & P(V):D^{(*)}=\left(D^{(*)0}~D^{(*)+}~D^{(*)+}_{s}\right), ~~{B}^{(*)}=\left(B^{(*)-}~\bar{B}^{(*)0}~\bar{B}^{(*)0}_{s}\right).\nonumber
 \end{eqnarray}
And the trace in Eqs.~(\ref{eq:LAPBB}-\ref{eq:LAVBB}) need be done in flavor space.
The strong couplings of ${\cal B}_{QQ^{\prime}}{\cal B}_{Q}P$ and ${\cal B}_{QQ^{\prime}}{\cal B}_{Q}V$
can be defined as follows. 
\begin{eqnarray}
 &  & \langle{\cal B}_{Q}(p,s_{z})P(q,s)|{\cal L}_{P{\cal B}_{Q}{\cal B}_{QQ^{\prime}}}|{\cal B}_{QQ^{\prime}}(p^{\prime},s_{z}^{\prime})\rangle=g_{P{\cal B}_{Q}{\cal B}_{QQ^{\prime}}}\bar{u}(p,s_{z})i\gamma_{5}u(p^{\prime},s_{z}^{\prime}),\label{eq:strongPBBc}\\
 &  & \langle{\cal B}_{Q}(p,s_{z})V(q,s)|{\cal L}_{V{\cal B}_{Q}{\cal B}_{QQ^{\prime}}}|{\cal B}_{QQ^{\prime}}(p^{\prime},s_{z}^{\prime})\rangle\nonumber \\
 &  & =f_{1V{\cal B}_{Q}{\cal B}_{QQ^{\prime}}}\bar{u}(p,s_{z})\slashed\varepsilon(q,s)u(p^{\prime},s_{z}^{\prime})+\frac{f_{2V{\cal B}_{Q}{\cal B}_{QQ^{\prime}}}}{m_{{{\cal B}_{Q}}}+m_{{{\cal B}_{QQ^{\prime}}}}}\bar{u}(p,s_{z})\sigma_{\mu\nu}(-iq^{\mu})\varepsilon^{\nu}(q,s)u(p^{\prime},s_{z}^{\prime}),\label{eq:strongVBBc}
\end{eqnarray}
where the momentum of doubly heavy baryon is $p^{\prime\mu}=p^{\mu}+q^{\mu}$. In this work, the relevant elements used to define the decay constants of heavy meson $P~(V)$ and ${\cal B}_{QQ^{\prime}}$ baryon are given as following, 
\begin{eqnarray}
 &  & \langle{\cal B}_{QQ^{\prime}}(p^{\prime},s_{z}^{\prime})|j_{{\cal B}_{QQ^{\prime}}}|0\rangle=f_{{\cal B}_{QQ^{\prime}}}\bar{u}(p^{\prime},s_{z}^{\prime}),\label{eq:decayconstantXicc}\\
 &  & \langle0|j_{5}|P\rangle=f_{P}m_{P}^{2},\label{eq:decayconstantP}\\
 &  & \langle0|j_{\mu}|V\rangle=f_{V}m_{V}\varepsilon_{\mu}^{*},\label{eq:decayconstantV}
\end{eqnarray}
where $j_{5}$ and $j_{\mu}$ represents the pseudoscalar and vector current respectively. After inserting the complete sets of doubly heavy baryon and heavy meson states into Eqs.~(\ref{eq:correlation5})-(\ref{eq:correlationmu}), and performing the
Fourier integration over $x$, we get
\begin{eqnarray}
  &  & \Pi_{\mu}(p,q)=\frac{\langle 0|{j_{\mu}|V(q)\rangle\langle{\cal B}_{Q}(p)V(q)|{\cal B}_{QQ^{\prime}}(p+q)\rangle\langle{\cal B}_{QQ^{\prime}}(p+q)|\bar{j}_{{\cal B}_{QQ^{\prime}}}}|0\rangle}{[(p+q)^{2}-m_{{\cal B}_{QQ^{\prime}}}^{2}][q^{2}-m_{P}^{2}]}+\cdots,\label{eq:correlationha5}\\
  &  & \Pi_{5}(p,q)=\frac{\langle 0|{j_{5}|P(q)\rangle\langle{\cal B}_{Q}(p)P(q)|{\cal B}_{QQ^{\prime}}(p+q)\rangle\langle{\cal B}_{QQ^{\prime}}(p+q)|\bar{j}_{{\cal B}_{QQ^{\prime}}}}|0\rangle}{[(p+q)^{2}-m_{{\cal B}_{QQ^{\prime}}}^{2}][q^{2}-m_{V}^{2}]}+\cdots.\label{eq:correlationhav}
 \end{eqnarray}
The final expression for the hadron level of the correlation function of strong coupling ${\cal B}_{QQ^{\prime}}{\cal B}_{Q}P_{Q^{\prime}}$ is
obtained by inserting Eqs. (\ref{eq:strongPBBc}-\ref{eq:decayconstantV}) into Eqs. (\ref{eq:correlationha5}-\ref{eq:correlationhav}) and summing over spins:
\begin{eqnarray}
 &  & \Pi_{5}^{H}(p,q)=-ig_{P{\cal B}_{Q}{\cal B}_{QQ^{\prime}}}f_{P}m_{P}^{2}f_{{\cal B}_{QQ^{\prime}}}\frac{\bar{u}_{{\cal B}_{Q}}[(m_{{\cal B}_{Q}}-m_{{\cal B}_{QQ^{\prime}}})\gamma_{5}+\slashed{q}\gamma_{5}]}{[(p+q)^{2}-m_{{\cal B}_{QQ^{\prime}}}^{2}][q^{2}-m_{P}^{2}]}+\cdots.\label{eq:hadronpbbf}
\end{eqnarray}
Similarly, the ones of ${\cal B}_{QQ^{\prime}}{\cal B}_{Q}V_{Q^{\prime}}$ can be derived as
\begin{eqnarray}
  \Pi_{\mu}^{H}(p,q)
 & = & \frac{-if_{V}m_{V}f_{{\cal B}_{QQ^{\prime}}}}{[(p+q)^{2}-m_{{\cal B}_{QQ^{\prime}}}^{2}][q^{2}-m_{V}^{2}]}[f_{1V{\cal B}_{Q}{\cal B}_{QQ^{\prime}}}\bar{u}_{{\cal B}_{Q}}\slashed\varepsilon(q,s)(\slashed p+\slashed q+m_{{\cal B}_{QQ^{\prime}}})\varepsilon_{\mu}^{*}(D^{*}) \nonumber\\
 &  &+\frac{f_{2V{\cal B}_{Q}{\cal B}_{QQ^{\prime}}}}{m_{{{\cal B}_{Q}}}+m_{{{\cal B}_{QQ^{\prime}}}}}\bar{u}_{{\cal B}_{Q}}\sigma_{\rho\sigma}(-iq^{\rho})\varepsilon^{\sigma}(q,s)(\slashed p+\slashed q+m_{{\cal B}_{QQ^{\prime}}})\varepsilon_{\mu}^{*}(D^{*}) ]+\cdots\nonumber\\
 &=&\Pi_{\mu}^{1H}(p,q)+\Pi_{\mu}^{2H}(p,q)+\cdots,\label{eq:hadronvbb}
\end{eqnarray}
where the two parts of the correlation function can be written as follows,
\begin{eqnarray}
  \Pi_{\mu}^{1H}(p,q)
  &=& \frac{-if_{V}m_{V}f_{{\cal B}_{QQ^{\prime}}}f_{1V{\cal B}_{Q}{\cal B}_{QQ^{\prime}}}}{[(p+q)^{2}-m_{{\cal B}_{QQ^{\prime}}}^{2}][q^{2}-m_{V}^{2}]}
 \bar{u}_{{\cal B}_{Q}}[-2m_{{\cal B}_{Q}}v_{\mu}-(m_{{\cal B}_{QQ^{\prime}}}-m_{{\cal B}_{Q}})\gamma_{\mu}-\gamma_{\mu}\slashed q\nonumber\\
 &&+(\frac{m_{{\cal B}_{QQ^{\prime}}}^2-m_{{\cal B}_{Q}}^2}{m_{V}^{2}})q^{\mu}+\frac{q_{\mu}\slashed q}{m_{V}^{2}}(m_{{\cal B}_{QQ^{\prime}}}-m_{{\cal B}_{Q}})],\label{eq:hvbbmu1}\\
  \Pi_{\mu}^{2H}(p,q)&=& 
  \frac{-if_{V}m_{V}f_{{\cal B}_{QQ^{\prime}}}f_{2V{\cal B}_{Q}{\cal B}_{QQ^{\prime}}}}{[(p+q)^{2}-m_{{\cal B}_{QQ^{\prime}}}^{2}][q^{2}-m_{V}^{2}]}\bar{u}_{{\cal B}_{Q}}[(m_{{\cal B}_{QQ^{\prime}}}-m_{{\cal B}_{Q}})\gamma^{\mu}+\gamma^{\mu}\slashed q-q^{\mu}\nonumber\\
 && -\frac{2m_{{\cal B}_{Q}}v_{\mu}\slashed q}{(m_{{\cal B}_{Q}}+m_{{\cal B}_{QQ^{\prime}}})}-\frac{q^{\mu}\slashed q}{(m_{{\cal B}_{Q}}+m_{{\cal B}_{QQ^{\prime}}})}].\label{eq:hvbbmu2}
\end{eqnarray}
Due to the spin summation, the final expression includes several $\gamma$-matrix structures. In this work, we will select the structure $\slashed q \gamma_{5}$ to perform the analysis of the strong coupling $g_{P{\cal B}_{Q}{\cal B}_{QQ^{\prime}}}$ as shown in Eq.~(\ref{eq:hadronpbbf}), and the structure $v_{\mu}(v_{\mu}\slashed q)$ to derive the coupling $f_{1(2)V{\cal B}_{Q}{\cal B}_{QQ^{\prime}}}$, as shown in Eqs.~(\ref{eq:hvbbmu1})-(\ref{eq:hvbbmu2}).
\subsection{QCD level from the LCDAs of singly heavy baryon anti-triplet}
\label{subsec:qcdlevel3}
The main theoretical task is the calculation of the correlation function Eqs.~(\ref{eq:correlation5})-(\ref{eq:correlationmu}) in QCD level. Here, we introduce the LCDAs for singly heavy baryons in light-cone coordinates with the spin-parity $j^{P}=0^{+}$ diquark. 
At the light cone $x^{2}=0$, the LCDAs of singly heavy baryon can be given as follows: 
\begin{align}
&\epsilon_{abc}\langle{\cal B}_{Q}(v)|\bar{q}_{1k}^{a}(t_{1})\bar{q}_{2i}^{b}(t_{2})\bar{Q}_{\gamma}^{c}(0)|0\rangle \nonumber\\
& =\frac{1}{8}v_{+}\psi^{n*}(t_{1},t_{2})f^{(1)}\bar{u}_{\gamma}(C^{-1}\gamma_{5}\bar{\slashed n})_{ki}
  -\frac{1}{8}\psi^{n\bar{n}*}(t_{1},t_{2})f^{(2)}\bar{u}_{\gamma}(C^{-1}\gamma_{5}i\sigma^{\mu\nu})_{ki}\bar{n}_{\mu}n_{\nu}\nonumber\\
 & +\frac{1}{4}\psi^{1*}(t_{1},t_{2})f^{(2)}\bar{u}_{\gamma}(C^{-1}\gamma_{5})_{ki}
  +\frac{1}{8v_{+}}\psi^{\bar{n}*}(t_{1},t_{2})f^{(1)}\bar{u}_{\gamma}(C^{-1}\gamma_{5}\slashed n)_{kl},\label{eq:lcda}
\end{align}
where $a$, $b$, $c$ are the color indicators and $v$ denotes the velocity of the singly heavy baryon.
The light-cone vectors $n$ should be expressed by the Lorentz covariant form in terms of $x$ and $v$:
\begin{align}
n_{\mu} & =\frac{1}{v\cdot x}x_{\mu},\ \ \ \bar{n}_{\mu}=2v_{\mu}-\frac{1}{v\cdot x}x_{\mu},
\label{eq:nmu}
\end{align}
The Fourier-transformed form of the LCDAs can be given as, 
\begin{align}
\psi(x_{1},x_{2})=\int_{0}^{\infty}d\omega_{1}d\omega_{2}e^{-i\omega_{1}t_{1}}e^{-i\omega_{2}t_{2}}\psi(\omega_{1},\omega_{2}).
\end{align}
Here the momentum of the two light quarks $\omega_{1}$ and $\omega_{2}$ are along the light-cone direction. The total magnitude of diquark momentum is 
  $\omega=\omega_{1}+\omega_{2}$. Note that $x_{1}=t_{1}n$ , $x_{2}=t_{2}n$
  \begin{eqnarray}
  \psi(t_{1},t_{2}) &=&\int_{0}^{\infty}d\omega d\omega_{2}e^{-i\omega t_{1}}e^{-i\omega_{2}(t_{2}-t_{1})}\psi(\omega_{1},\omega_{2}), \\
  \psi(0,t_{2})&=&\int_{0}^{\infty}d\omega\omega\int_{0}^{1}due^{-i\bar{u}\omega v\cdot x_{2}}\psi(\omega,u),
  \end{eqnarray}
  where $t_{i}=v\cdot x_{i}$ and $\omega_2=(1-u)\omega=\bar u \omega$.  
And the LCDAs in Eq.~(\ref{eq:lcda}) can be renamed into different twistes:
\begin{align*}
\psi^{n}(\omega,u)\rightarrow\psi_{2}(\omega,u),\ \ \psi^{n\bar{n}}(\omega,u)\rightarrow\psi_{3a}(\omega,u),\ \ \psi^{1}(\omega,u)\rightarrow\psi_{3s}(\omega,u),\ \ \psi^{\bar{n}}(\omega,u)\rightarrow\psi_{4}(\omega,u).
\end{align*}
The correlation functions with the pseudoscalar and vector current can be written as follows,
\begin{eqnarray}
&&\Pi_{5}^{QCD}(p,q) \nonumber \\
&& =  i\int d^{4}xe^{iq\cdot x}\langle{\cal B}_{Q}(p)|T\{j_{5}(x)\bar{j}_{{\cal B}_{QQ^{\prime}}}(0)\}|0\rangle\nonumber \\
 &  &= -2i\int d^{4}xe^{iq\cdot x}\epsilon_{abc}\{\langle{\cal B}_{Q}(p)|\bar{q}_{a}(0)_{k}\bar{q}_{b}(x)_{i}\bar{Q}_{c}(0)_{m}|0\rangle(i\gamma_{5}S^{Q}(x)\gamma^{\nu}C)_{im}(\gamma_{\nu}\gamma_{5})_{kl}\},\label{eq:qcdp}\\
 &&\Pi_{\mu}^{QCD}(p,q) \nonumber \\
 && = i\int d^{4}xe^{iq\cdot x}\langle{\cal B}_{Q}(p)|T\{j_{\mu}(x)\bar{j}_{{\cal B}_{QQ^{\prime}}}(0)\}|0\rangle\nonumber\\ 
 && =  -2i\int d^{4}xe^{iq\cdot x}\epsilon_{abc}\{\langle{\cal B}_{Q}(p)|\bar{q}_{a}(0)_{k}\bar{q}_{b}(x)_{i}\bar{Q}_{c}(0)_{m}|0\rangle(\gamma_{\mu}S^{Q}(x)\gamma^{\nu}C)_{im}(\gamma_{\nu}\gamma_{5})_{kl}\}.\label{eq:qcdv}
\end{eqnarray}
here $\Pi_{\mu(5)}^{QCD}(p,q)$ represents the QCD level correlation function with the pseudoscalar (vector) current. In the following, we discuss QCD level calculation of the strong couplings in two parts. 
\begin{itemize}
  \item Substitute the expression Eqs.~(\ref{eq:lcda}-\ref{eq:nmu}) into Eq.~(\ref{eq:qcdp}), the correlator function of the pseudoscalar current can be obtained as: 
\begin{align*}
\Pi_{5}^{QCD}(p,q) & =-\frac{i}{4}\int d^{4}x\int_{0}^{\infty}d\omega\omega\int_{0}^{1}due^{i(q+\bar{u}\omega v)\cdot x}\\
 & \times\{v_{+}\psi_{2}(\omega,u)f^{(1)}\bar{u}_{{\cal B}_{Q}}[\gamma^{\nu}CS^{Q}(x)^{T}C^{T}(-i\gamma_{5})(2\slashed v-\frac{\slashed x}{v\cdot x})\gamma_{\nu}]\\
 & -\psi_{3a}(\omega,u)f^{(2)}\bar{u}_{{\cal B}_{Q}}[\gamma^{\nu}CS^{Q}(x)^{T}C^{T}(-i\gamma_{5})i\sigma^{\alpha\beta}\gamma_{\nu}](2v_{\alpha}-\frac{x_{\alpha}}{v\cdot x})\frac{x_{\beta}}{v\cdot x}\\
 & -2\psi_{3s}(\omega,u)f^{(2)}\bar{u}_{{\cal B}_{Q}}[\gamma^{\nu}CS^{Q}(x)^{T}C^{T}(-i\gamma_{5})\gamma_{\nu}]\\
 & +\frac{1}{v_{+}}\psi_{4}(\omega,u)f^{(1)}\bar{u}_{{\cal B}_{Q}}[\gamma^{\nu}CS^{Q}(x)^{T}C^{T}(-i\gamma_{5})\frac{\slashed x}{v\cdot x}\gamma_{\nu}]\},
\end{align*}
in which $C$ is the charge conjugation operator and $C^T = C^{-1} = C^{\dagger}= -C$.
In the Euclidean region, both virtualities $p^2$ and $(p+q)^2$ are
negative and large, so that the heavy quark is sufficiently far off-shell. As a first
approximation, the free heavy quark propagator can be given as follows,
\begin{align}
CS^{Q}(x)^{T}C^{T}=-\int\frac{d^{4}k}{(2\pi)^{4}}\frac{i(\slashed k-m_{Q})}{k^{2}-m_{Q}^{2}+i\epsilon}e^{-ik\cdot x}.\label{eq:propagator}
\end{align}
Using the above Eq.~(\ref{eq:propagator}), the correlation function with the pseudoscalar current is expressed as a convolution by $u$ and $\omega$ on the QCD side and can be divided into two parts.
\begin{align}
  \Pi_{5}^{QCD}(p,q) & =\bar{u}_{{\cal B}_{Q}}[C_{\gamma_{5}}\gamma_{5}+C_{\slashed q\gamma_{5}}\slashed q\gamma_{5}],\label{eq:qcdfp}
  \end{align}
with the coeffections 
  \begin{align}
  C_{\gamma_{5}} & =i\int_{0}^{\infty}d\omega\int_{0}^{1}du\big\{\omega v_{+}\psi_{2}(\omega,u)f^{(1)}[{2\bar{u}\omega+2q\cdot v+m_{Q}}]\frac{1}{\Delta}\nonumber\\
   & +\bar{u}v_{+}f^{(1)}\tilde{\psi}_{2}(\omega,u)\big[{2(q+\bar{u}\omega v)^{2}-m_{Q}\bar{u}\omega-4m_{Q}^{2}}\big]\frac{1}{\Delta^{2}}\nonumber\\
   & -\bar{u}\tilde{\psi}_{3a}(\omega,u)[{(q^{2}+4\bar{u}\omega q\cdot v+3\bar{u}^{2}\omega^{2})-3m_{Q}^{2}}]\frac{1}{\Delta^{2}}\nonumber\\
   & -\omega{\psi}_{3s}(\omega,u)f^{(2)}[\bar{u}\omega+2m_{Q}]\frac{1}{\Delta}\nonumber\\
   & +\bar{u}\tilde{\psi}_{4}(\omega,u)f^{(1)}[{2(q+\bar{u}\omega v)^{2}-4m_{Q}^{2}-m_{Q}\bar{u}\omega}]\frac{1}{\Delta^{2}}\big\},
   \label{eq:cg5}\\
  C_{\slashed q\gamma_{5}} & =-i\int_{0}^{\infty}d\omega\int_{0}^{1}du\big\{\bar{u}v_{+}f^{(1)}\tilde{\psi}_{2}(\omega,u)\big[{+m_{Q}}\big]\frac{1}{\Delta^{2}}\nonumber\\
  & +\bar{u}\tilde{\psi}_{3a}(\omega,u){[2(q\cdot v+\bar{u}\omega)]}\frac{1}{\Delta^{2}}\nonumber\\
   & +\omega{\psi}_{3s}(\omega,u)f^{(2)}\frac{1}{\Delta}+\bar{u}\tilde{\psi}_{4}(\omega,u)f^{(1)}[{m_{Q}}]\frac{1}{\Delta^{2}}\big\}. \label{eq:cqsg5}
  \end{align}
Here $\Delta=(q+\bar{u}\omega v)^{2}-m_{Q}^{2}$, 
and $m_{Q}$ is the mass of the heavy quark $Q=b,~c$.
In order to extract the strong coupling constant $g_{P{\cal B}_{Q}{\cal B}_{QQ^{\prime}}}$, we choose the coeffections of Lorentz structure $\slashed q\gamma_{5}$.

\item Similarly, substitute Eqs.~(\ref{eq:lcda}-\ref{eq:nmu}) into Eq.~(\ref{eq:qcdv}), the correlator function of the vector current can be obtained as:
\begin{align}
  \Pi_{\mu}^{QCD}(p,q) & =-\frac{i}{4}\int d^{4}x\int_{0}^{\infty}d\omega\omega\int_{0}^{1}due^{i(q+\bar{u}\omega v)\cdot x}\nonumber\\
   & \times\{v_{+}\psi_{2}(\omega,u)f^{(1)}\bar{u}_{{\cal B}_{Q}}[\gamma^{\nu}CS^{Q}(x)^{T}C^{T}\gamma_{\mu}(2\slashed v-\frac{\slashed x}{v\cdot x})\gamma_{\nu}] \nonumber\\
   & -\psi_{3a}(\omega,u)f^{(2)}\bar{u}_{{\cal B}_{Q}}[\gamma^{\nu}CS^{Q}(x)^{T}C^{T}\gamma_{\mu}i\sigma^{\alpha\beta}\gamma_{\nu}](2v_{\alpha}-\frac{x_{\alpha}}{v\cdot x})\frac{x_{\beta}}{v\cdot x} \nonumber\\
   & -2\psi_{3s}(\omega,u)f^{(2)}\bar{u}_{{\cal B}_{Q}}[\gamma^{\nu}CS^{Q}(x)^{T}C^{T}\gamma_{\mu}\gamma_{\nu}]\nonumber\\
   & +\frac{1}{v_{+}}\psi_{4}(\omega,u)f^{(1)}\bar{u}_{{\cal B}_{Q}}[\gamma^{\nu}CS^{Q}(x)^{T}C^{T}\gamma_{\mu}\frac{\slashed x}{v\cdot x}\gamma_{\nu}]\}.\label{eq:correlationmun}
  \end{align}
  Using the Eq.~(\ref{eq:propagator}), the correlation function with the vector current on the QCD side, can be expressed with six Lorentz structures.
\begin{align}
    &\Pi_{\mu}^{QCD}(p,q) =\bar{u}_{{\cal B}_{Q}}[C_{v_{\mu}}v_{\mu}+C_{\gamma_{\mu}}\gamma_{\mu}+C_{q_{\mu}}q_{\mu}+C_{v_{\mu}\slashed q}v_{\mu}\slashed q+C_{\gamma_{\mu}\slashed q}\gamma_{\mu}\slashed q+C_{q_{\mu}\slashed q}q_{\mu}\slashed q],\label{eq:qcdfv}
\end{align}
with the corresponding coeffections
\begin{align}
C_{v_{\mu}} & =\int_{0}^{\infty}d\omega\int_{0}^{1}du\big\{\omega v_{+}\psi_{2}(\omega,u)f^{(1)}[2(\bar{u}\omega+m_{Q})]\frac{1}{\Delta}\nonumber\\
&-\bar{u}f^{(1)}\tilde{\psi}_{2}(\omega,u)\left[2\bar{u}\omega(m_{Q}+\bar{u}\omega)\right]\frac{1}{\Delta^{2}} \nonumber\\
&-2\bar{u}\tilde{\psi}_{3a}(\omega,u)f^{(2)}[\frac{1}{\Delta}+2(\bar{u}^{2}\omega^{2}+q\cdot v\bar{u}\omega-m_{Q}^{2})\frac{1}{\Delta^{2}}]\nonumber\\
 &+2\omega{\psi}_{3s}(\omega,u)f^{(2)}[\bar{u}\omega]\frac{1}{\Delta} +\bar{u}f^{(1)}\tilde{\psi}_{4}(\omega,u)\left[2\bar{u}\omega(m_{Q}+\bar{u}\omega)\right]\frac{1}{\Delta^{2}}\big\},\label{eq:cvm}\\
C_{\gamma_{\mu}} & =\int_{0}^{\infty}d\omega\int_{0}^{1}du\big\{-\bar{u}\omega^{2}v_{+}\psi_{2}(\omega,u)f^{(1)}\frac{1}{\Delta}+2\bar{u}\tilde{\psi}_{3a}(\omega,u)f^{(2)}m_{Q}q\cdot v\frac{1}{\Delta^{2}}\nonumber\\
&+\bar{u}f^{(1)}\tilde{\psi}_{2}(\omega,u)m_{Q}^{2}\frac{1}{\Delta^{2}} +\omega{\psi}_{3s}(\omega,u)f^{(2)}m_{Q}\frac{1}{\Delta}-\bar{u}f^{(1)}\tilde{\psi}_{4}(\omega,u)m_{Q}^{2}\frac{1}{\Delta^{2}}\big\},\label{eq:cgm}\\
C_{q_{\mu}} & =\int_{0}^{\infty}d\omega\int_{0}^{1}du\big\{-2f^{(1)}\tilde{\psi}_{2}(\omega,u)(\bar{u}m_{Q}+\bar{u}^{2}\omega)\frac{1}{\Delta^{2}}\nonumber\\
&+4\tilde{\psi}_{3a}(\omega,u)f^{(2)}(\bar{u}^{2}\omega+\bar{u}q\cdot v+\bar{u}m_{Q})\frac{1}{\Delta^{2}}\nonumber\\
 & +2\omega{\psi}_{3s}(\omega,u)f^{(2)}\frac{1}{\Delta}+2f^{(1)}\tilde{\psi}_{4}(\omega,u)(\bar{u}m_{Q}+\bar{u}^{2}\omega)\frac{1}{\Delta^{2}}\big\},\label{eq:cqm}\\
C_{v_{\mu}\slashed q} & =\int_{0}^{\infty}d\omega\int_{0}^{1}du\big\{\bar{u}f^{(1)}\tilde{\psi}_{2}(\omega,u)\left[-2\bar{u}\omega\right]\frac{1}{\Delta^{2}}+\bar{u}f^{(1)}\tilde{\psi}_{4}(\omega,u)\left[2\bar{u}\omega\right]\frac{1}{\Delta^{2}}\big\},\label{eq:cvmqs}\\
C_{\gamma_{\mu}\slashed q} & =\int_{0}^{\infty}d\omega\int_{0}^{1}du\big\{\omega v_{+}\psi_{2}(\omega,u)f^{(1)}\frac{1}{\Delta}-2\bar{u}\tilde{\psi}_{3a}(\omega,u)f^{(2)}m_{Q}\frac{1}{\Delta^{2}}\big\},\label{eq:cgmqs}\\
C_{q_{\mu}\slashed q} & =\int_{0}^{\infty}d\omega\int_{0}^{1}du\big\{-2\bar{u}f^{(1)}\tilde{\psi}_{2}(\omega,u)\frac{1}{\Delta^{2}}+2\bar{u}f^{(1)}\tilde{\psi}_{4}(\omega,u)\frac{1}{\Delta^{2}}\big\}.\label{eq:cqmqs}
\end{align}
There exist several ways to get a particular result for the strong coupling constant $f_{V{\cal B}_{Q}{\cal B}_{QQ^{\prime}}}$. Hence we are free to choose any one of these ways. In this work, we have utilized our freedom to choose the coeffections corresponding to the Lorentz structures $v_{\mu}$ and $v_{\mu}\slashed q$.
\end{itemize}
\subsection{Quark-Hadron Duality}
\label{subsec:quark hadronquality}
On the hadron level, to remove the contributions from higher states and the continuum, we utilize the double Borel transformation concerning the square of the momenta $p^{\prime 2}=(p+q)^2$ and $ q^2$, which leads to 
\begin{eqnarray}\label{eq:correlationhag}
  &  & {\cal B}_{p^{\prime}}(M_1^2){\cal B}_{q}(M_2^2)\Pi_{5}^{H}(p,q)=-ig_{P{\cal B}_{Q}{\cal B}_{QQ^{\prime}}}f_{P}m_{P}^{2}f_{{\cal B}_{QQ^{\prime}}}e^{-\frac{m_{P}^2}{M_2^2}-\frac{m_{{\cal B}_{QQ^{\prime}}}^2}{M_1^2}}{\bar{u}_{{\cal B}_{Q}}[\slashed q\gamma_{5}]}+\cdots.\label{eq:hadronpbb}\\
  & &{\cal B}_{p^{\prime}}(M_1^2){\cal B}_{q}(M_2^2)\Pi_{\mu}^{H}(p,q)
  = -if_{V}m_{V}f_{{\cal B}_{QQ^{\prime}}}e^{-\frac{m_{V}^2}{M_2^2}-\frac{m_{{\cal B}_{QQ^{\prime}}}^2}{M_1^2}}\nonumber\\
  &&\qquad\qquad\qquad\qquad\times\bar{u}_{{\cal B}_{Q}}[f_{1V{\cal B}_{Q}{\cal B}_{QQ^{\prime}}}
 (-2m_{{\cal B}_{Q}}v_{\mu})-f_{2V{\cal B}_{Q}{\cal B}_{QQ^{\prime}}}\frac{2m_{{\cal B}_{Q}}v_{\mu}\slashed q}{(m_{{\cal B}_{Q}}+m_{{\cal B}_{QQ^{\prime}}})}]+\cdots.
 \end{eqnarray}
 The Borel parameters $M^2_1$ and $M^2_2$ are associated with the square momenta $p^{\prime 2}$ and $q^2$, respectively. 
 Given that the masses of the initial doubly heavy baryon are approximately twice those of the final heavy meson, the Borel parameters are chosen to reflect this ratio.
 According to Eqs.~(\ref{eq:qcdfp}) and (\ref{eq:qcdfv}), we choose the corresponding structures $\slashed q\gamma_{5}$, $v_{\mu}$ and $v_{\mu}\slashed q$ on the QCD side as well.
Therefore, one can write the final result using the Quark-Hadron Duality,
   \begin{align}
    &g_{P{\cal B}_{Q}{\cal B}_{QQ^{\prime}}}=\frac{i}{f_{P}m_{P}^2
    f_{{\cal B}_{QQ^{\prime}}}}{\cal B}_{p^{\prime}}(M_1^2){\cal B}_{q}(M_2^2)
    C_{\slashed q\gamma_5}e^{\frac{m_{P}^{2}}{M_{2}^2}+\frac{m_{{\cal B}_{QQ^{\prime}}}^{2}}{M_{1}^2}},\label{eq:hqgpbb}\\
      & {f_{1V{\cal B}_{Q}{\cal B}_{QQ^{\prime}}}}=\frac{-i}{2m_{{\cal B}_{Q}}f_{V}m_{V}f_{{\cal B}_{QQ^{\prime}}}}{\cal B}_{p^{\prime}}(M_1^2){\cal B}_{q}(M_2^2)C_{v_{\mu}}e^{\frac{m_{V}^{2}}{M_{2}^2}+\frac{m_{{\cal B}_{QQ^{\prime}}}^{2}}{M_{1}^2}},\label{eq:hqf1vbb}\\
      & {f_{2V{\cal B}_{Q}{\cal B}_{QQ^{\prime}}}}=\frac{-i(m_{{\cal B}_{Q}}+m_{{\cal B}_{QQ^{\prime}}})}{2m_{{\cal B}_{Q}}f_{V}m_{V}f_{{\cal B}_{QQ^{\prime}}}}{\cal B}_{p^{\prime}}(M_1^2){\cal B}_{q}(M_2^2)C_{v_{\mu}\slashed q}e^{\frac{m_{V}^{2}}{M_{2}^2}+\frac{m_{{\cal B}_{QQ^{\prime}}}^{2}}{M_{1}^2}}.\label{eq:hqf2vbb}
   \end{align}
   Next, we need to apply a double Borel transformation to the $1/\Delta^n$ in Eqs.~(\ref{eq:cqsg5}), (\ref{eq:cvm}) and (\ref{eq:cvmqs}). Before this transformation, it is necessary to represent the $1/\Delta^n$ in the form of double spectral integrals, as demonstrated in Ref.~\cite{Belyaev:1994zk}.
   \begin{align*}
    &\int^{\infty}_{0}d\omega\int^{1}_{0}du\frac{\psi(\omega,u)}{[(q+\bar{u}\omega v)^{2}-m_{Q}^{2}]^n}=\int^{\infty}_{0}d\omega\int^{1}_{0}du\frac{-\psi(\omega,u)}{[\beta-(1-\alpha)q^2-\alpha(p+q)^2]^n}\\
    &=\int^{\infty}_{0}d\omega\frac{m_{B_{Q}}}{\omega}\int^{\frac{\omega}{m_{B_{Q}}}}_{0}d\alpha\frac{-\psi(\omega,1-\frac{m_{B_{Q}}}{\omega}\alpha)}{[\beta-(1-\alpha)q^2-\alpha(p+q)^2]^n},
    \end{align*}
  with
    \begin{align*}
    &\alpha=\frac{\bar{u}\omega}{m_{{\cal B}_{Q}}},~\beta=m_{Q}^2+\bar{u}\omega m_{{\cal B}_{Q}}-(\bar{u}\omega)^2=m_{Q}^2+\alpha(1-\alpha) m_{{\cal B}_{Q}}^2.
  \end{align*}
  Since the integration limit for $\alpha$ is $[0,1]$, the upper limit of the $\omega$ integral should be $m_{{\cal B}_{Q}}$.
  It is sufficient to identify double spectral representations for the master integrals of the form $\int d\alpha \alpha^k {\Delta}^{-n}$ where $k \geq 0$ and $n = 1, 2$~\cite{Khodjamirian:2011jp}.
For $n=1$, we obtain
\begin{eqnarray}
 \int_0^1 d \alpha   { \alpha^k \over  {\Delta} } =  {1 \over 2 \pi} \sum_{j=0}^{k}
\int_{(m_Q+m_{Q^{\prime}})^2}^{\infty}   {d s \over s - (p+q)^2} \int\limits_{t_1(s)}^{t_2(s)}
{d s^{\prime} \over s^{\prime}- q^2 }(-1)^{k+j/2}
[1+(-1)^j ] &&  \nonumber \\
\times {1\over (2 m_{{\cal B}_{Q}}^2)^k  } C_k^j     (s - s^{\prime}-m_{{\cal B}_{Q}}^2)^{k-j}
[(s^{\prime}-t_1)(t_2-s^{\prime})]^{ j-1 \over 2 }\,,
\end{eqnarray}
where  $C_k^j$ are the binomial coefficients and the integration limits are as follows:
\begin{eqnarray}
t_{1,2} (s)= (s+m_{{\cal B}_{Q}}^2)  \mp 2 m_{{\cal B}_{Q}} \sqrt{s-(m_Q+m_{Q^{\prime}})^2} \,.
\label{eq:intlim}
\end{eqnarray}
The double spectral representations for the master integrals with $n=2$, need to be completed ahead of Borel transformation,
\begin{eqnarray}
{\frac{1}{[(q+\bar{u}\omega v)^2-m_{Q}^2]^n}}\to\frac{1}{(n-1)!}\left(\frac{\partial}{\partial \Omega}\right)^{(n-1)}{\frac{1}{(q+\bar{u}\omega v)^2-\Omega}}\Big|_{\Omega=m_Q^2}.
\end{eqnarray}
You can also calculate it directly using the formula provided below,
\begin{eqnarray}
  \int_0^1 d \alpha   { \alpha^k \over \Delta^2 } & =  &  - {1 \over  \pi}\int_{(m_Q+m_{Q^{\prime}})^2}^{\infty}   {d s \over s -
 (p+q)^2}   \int_{t_1(s)}^{t_2(s)}  {d s^{\prime} \over s^{\prime}-
 q^2} \bigg \{  \sum _{j=2}^{k} \, (-1)^{k+1+j/2} \,\,  {1
 +(-1)^j \over 2}
 \nonumber \\
 &&  \times \frac{j-1}{ (2 m_N^2)^{k-1}} \, C_k^j \,
 [\bar{s}(s^{\prime})]^{k-j} \, [\kappa(s^{\prime},t_1, t_2)]^{ j-3
 \over 2 }  \theta(k-2)  \nonumber \\
 &&   + {(-1)^k \over (2 m_N^2)^{k-1} } {[\bar{s}(s^{\prime})]^k
 \over [\kappa(s^{\prime},t_1,
 t_2)]^{3/2} }   \nonumber \\
 && -  { (-1)^k \over (2 m_N^2)^{k-1} }  \bigg [   \bigg ({[
 \bar{s}(t_1)]^k \over t_2 -t_1 } \delta(s^{\prime}-t_1) X_1(t_1,
 t_2) \bigg )  -  \bigg ( t_1 \leftrightarrow t_2  \bigg )  \bigg ] \nonumber \\
  &&-\pi \delta((m_Q+m_{Q^{\prime}})^2 - s) \delta((m_Q+m_{Q^{\prime}})^2+m_{{\cal B}_{Q}}^2 - s^{\prime}) \bigg \}\,+
  ...\,,
 \end{eqnarray}
 here, the ellipses denote terms that become insignificant after undergoing a double Borel transformation and are deemed inessential.
 $t_{1,2}$ are the functions of $s$ defined in Eq.~(\ref{eq:intlim}), $\theta(k-a)= 1(0)$ at $k\geq a$($k<a$),
 $\bar{s}(y)=s-y-m_{{\cal B}_{Q}}^2$ and  $\kappa(a, b, c)=(a- b)(c-a)$. The auxiliary function employed in the above equation is delineated below:
 \begin{eqnarray}
 X_1(a, b) &=& \int_{a}^{b} {d \sigma  \over
 [\kappa(\sigma,a, b)]^{3/2} } (b -\sigma).
 \end{eqnarray}
 And for the terms including $\frac{q\cdot v}{\Delta^{2}}\tilde{\psi}(\omega,u)$, we first transform them as 
\begin{align}
  &\int du\int d\omega\frac{q\cdot v}{\Delta^{2}}\tilde{\psi}(\omega,u)=\int du\int d\omega\tilde{\psi}(\omega,u)\frac{-1}{2\bar{u}}[\frac{d}{d\omega}(\frac{1}{\Delta})+\frac{\bar{u}^{2}\omega}{\Delta^{2}}]\nonumber\\
  & =\int du\int d\omega\frac{-1}{2\bar{u}}\omega({\psi}(\omega,u))[(\frac{1}{\Delta})]-\frac{1}{2}\int du\int d\omega\tilde{\psi}(\omega,u)[\frac{\bar{u}\omega}{\Delta^{2}}].
   \end{align}
   With these integrals, one can easily write down the double dispersion representation for all integrals in Eqs.~(\ref{eq:hqgpbb}-\ref{eq:hqf2vbb}), where the numerators are polynomials in $\alpha$ and linear in $(p+q)^2$ and $q^2$.
   To facilitate the substitution of subsequent variables, they can be replaced with $s$ and $s^{\prime}$, respectively.

The OPE results are congruent with the double hadronic dispersion relations.
By employing the quark-hadron duality approximation for the hadronic spectral densities and implementing a double Borel transformation, $(p+q)^2\to M^2_{1}$, $q^2\to {M}^2_{2}$, we can derive LCSR for the three strong couplings,
\begin{eqnarray}
  &&g_{P{\cal B}_{Q}{\cal B}_{QQ^{\prime}}}=\frac{ie^{{m_{P}^{2}}/{M_{2}^2}+{m_{{\cal B}_{QQ^{\prime}}}^{2}}/{M_{1}^2}}}{f_{P}m_{P}^2
  f_{{\cal B}_{QQ^{\prime}}}}\frac{1}{\pi^2}\int_{s_{\rm min}}^{s_{\rm th}} ds\, e^{-s/M^2_{1}}\int_{t_1(s)}^{t_2(s)}  ds^{\prime}\,
  e^{-s^{\prime}/{M}^2_{2}}
  {\rm Im}_{s} \, {\rm Im}_{s^{\prime}}C_{\slashed q\gamma_5} (s, s^{\prime}),\label{eq:hqgpbbF}\\
    && {f_{1V{\cal B}_{Q}{\cal B}_{QQ^{\prime}}}}=\frac{-ie^{{m_{V}^{2}}/{M_{2}^2}+{m_{{\cal B}_{QQ^{\prime}}}^{2}}/{M_{1}^2}}}{2m_{{\cal B}_{Q}}f_{V}m_{V}f_{{\cal B}_{QQ^{\prime}}}}\frac{1}{\pi^2}\int_{s_{\rm min}}^{s_{\rm th}} ds\, e^{-s/M^2_{1}}\int_{t_1(s)}^{t_2(s)}  ds^{\prime}\,
    e^{-s^{\prime}/{M}^2_{2}}
    {\rm Im}_{s} \, {\rm Im}_{s^{\prime}}C_{v_{\mu}},\label{eq:hqf1vbbF}\\
    && {f_{2V{\cal B}_{Q}{\cal B}_{QQ^{\prime}}}}=\frac{-ie^{{m_{V}^{2}}/{M_{2}^2}+{m_{{\cal B}_{QQ^{\prime}}}^{2}}/{M_{1}^2}}(m_{{\cal B}_{Q}}+m_{{\cal B}_{QQ^{\prime}}})}{2m_{{\cal B}_{Q}}f_{V}m_{V}f_{{\cal B}_{QQ^{\prime}}}}\nonumber\\
    &&\quad\qquad\quad\qquad\times\frac{1}{\pi^2}\int_{s_{\rm min}}^{s_{\rm th}} ds\, e^{-s/M^2_{1}}\int_{t_1(s)}^{t_2(s)}  ds^{\prime}\,
    e^{-s^{\prime}/{M}^2_{2}}
    {\rm Im}_{s} \, {\rm Im}_{s^{\prime}}C_{v_{\mu}\slashed q}.\label{eq:hqf2vbbF}
\end{eqnarray}
Here, the integral limits of $s$ are noted with ${s_{\rm min}}=(m_{Q}+m_{Q^{\prime}})^2$ and the continus threshood ${s_{\rm th}}$. 
And the corresponding values of the Borel parameter \(M^2_{1,2}\) and continuum threshold ${s_{\rm th}}$ exhibit variations for diverse strong coupling types, as presented in Tab.~\ref{Tab:para_Mands}.
We can subsequently derive the ultimate expressions of the strong couplings. The subsequent section will present the derivation of the final numerical results for the strong couplings.
\section{Numerical results}
\label{sec:numericalresults}
In this section, we conduct a numerical analysis of the LCSR for strong coupling constants associated with heavy mesons and heavy baryons ${\cal B}_{QQ^{\prime}}{\cal B}_{Q}P_{Q^{\prime}}$ and ${\cal B}_{QQ^{\prime}}{\cal B}_{Q}V_{Q^{\prime}}$. Subsequently, we provide a comprehensive discussion of the results obtained. 
\subsection{Input paremeters}
The sum rules for the couplings ${\cal B}_{QQ^{\prime}}{\cal B}_{Q}P_{Q^{\prime}}$ and ${\cal B}_{QQ^{\prime}}{\cal B}_{Q}V_{Q^{\prime}}$ incorporate input parameters such as heavy quark masses, the mass and decay constant of the heavy meson, and singly and doubly heavy baryons. These parameters were either extracted from experimental data or computed using nonperturbative methods. 
In our numerical analysis, we utilize heavy quark masses of \(m_{c}(m_{c})=(1.35\pm0.10)\) GeV and \(m_{b}(m_{b})=(4.7\pm0.1)\) GeV in the $\overline{\rm MS} $ scheme~\cite{ParticleDataGroup:2024cfk}, while the masses of light quarks are neglected. The input parameters, namely mass and decay constant, for both doubly heavy baryons and singly heavy baryons are comprehensively listed in Tab.~\ref{tab:inputparameter}~\cite{Karliner:2014gca,Shah:2016vmd,Shah:2017liu,Kiselev:2001fw,Groote:1997yr,Wang:2010fq,Wang:2009cr,Agaev:2017jyt}.
Another set of important input parameters are the various twist wave functions of singly heavy baryons included in the LCDAs. The LCDAs used in this work can be found in Ref.~\cite{Ali:2012pn}, which only provides LCDAs of bottom baryons. But in heavy quark limit, one can safely apply these LCDAs for charm baryons~\cite{Shi:2019fph}. In this work both $\Lambda_{b}/\Xi_{b}$ and $\Lambda_{c}/\Xi_{c}$ are described by the same LCDAs given in Ref.~\cite{Ali:2012pn}, which are expressed as
  \begin{eqnarray}
   \psi_{2}(\omega,u) &=& \omega^2 \bar u u\sum_{n=0}^{2}
   \frac{a_{n}}{\epsilon_{n}^{4}}\frac{C_{n}^{3/2}(2u-1)}{|C_{n}^{3/2}|^2}e^{-\omega/\epsilon_{n}}\,,
  \nonumber\\
   \psi_{4}(\omega,u) &=& \sum_{n=0}^{2}
   \frac{a_{n}}{\epsilon_{n}^{2}}\frac{C_{n}^{1/2}(2u-1)}{|C_{n}^{1/2}|^2}e^{-\omega/\epsilon_{n}}\,,
  \nonumber\\
   \psi_{3\sigma,s}(\omega,u) &=& \frac{\omega}{2}\sum_{n=0}^{2}
   \frac{a_{n}}{\epsilon_{n}^{3}}\frac{C_{n}^{1/2}(2u-1)}{|C_{n}^{1/2}|^2}e^{-\omega/\epsilon_{n}}\,,
  \label{SR:pert}
  \end{eqnarray}
  with 
  \begin{eqnarray}
     |C_{n}^{\lambda}|^2 &=& \int_0^{1}[C_{n}^{\lambda}(2u-1)]^2\, \,,
  \end{eqnarray}
  where $C_{0}^{\lambda}(x)=1$, $C_{1}^{\lambda}(x)=2\lambda x$ and $C_{2}^{\lambda}(x)=2\lambda(1+\lambda) x^2-\lambda$.
  The parameters in Eq.~\eqref{SR:pert} are collected 
  in Tab.~\ref{modelparam}. The parameter $A$ is chosen to make sure that $\epsilon_i$s are non-negative. In this work, we carefully choose $A=0.24$ for $\Lambda_{b}/\Lambda_{c}$ and $A=0.10$ for $\Xi_{b}/\Xi_{c}$.
  \begin{table}
  \caption{
  Parameters for the LCDAs of $\Lambda_{b}$ and $\Xi_{b}$ in 
  Eqs.~(\ref{SR:pert}). A replacement $A \to 1-A$ is made for transversal LCDAs~\cite{Ali:2012pn}.
  2, $3\sigma$, 3s and 4 are twist notations. 
  }   
  \label{modelparam}
  \begin{center}
  \def\arraystretch{1.5}
  \begin{tabular}{c|c|p{2cm}p{2cm} p{2cm} p{2cm} p{2cm} p{2cm}}
  \hline\hline \multirow{6}{*}{\mbox{\Large $ \Lambda_b $}}    
    &twist& $a_0$ & $a_1$ & $a_2$ & $\varepsilon_0$[GeV] & $\varepsilon_1$[GeV] & $\varepsilon_2$[GeV]  \\ 
    \cline{2-8} 
    &$2      $  & $  1 $ & $ -  $ & $ \frac{6.4 A}{A+0.44}  $& $  \frac{1.4 A+0.6}{A+5.7} $& $ -  $ &$ \frac{0.32 A}{A-0.17}  $\\
    &$3 s    $  & $  1 $ & $ -  $ & $ \frac{0.12 A-0.08}{A-1.4}  $& $ \frac{0.56 A-0.77}{A-2.6}  $& $ -  $ &$ \frac{0.25 A-0.16}{A+0.41}  $\\
    &$3\sigma     $  & $ -  $ & $ 1  $ & $ -  $& $ -  $& $ \frac{  0.35 A -0.43   }{A-1.2 }  $ &$ -  $\\
    &$4      $  & $ 1  $ & $ -  $ & $ \frac{  -0.07 A - 0.05   }{A +0.34 }  $& $ \frac{  0.65 A+0.22   }{A+1 }  $& $ -  $ &$ \frac{  5.5 A+3.8   }{A +29 }  $\\
     \hline\hline \multirow{6}{*}{\mbox{\Large$ \Xi_b  $}}  
    &twist& $a_0$ & $a_1$ & $a_2$ & $\varepsilon_0$ & $\varepsilon_1$ & $\varepsilon_2$  \\ 
    \cline{2-8}
    &$2      $  & $1$ & $\frac{0.25 A+0.46}{A+0.68}$ & $\frac{6.6A+0.6}{A+0.68}$& $\frac{1.4 A+1}{A+6.7}$& $\frac{0.57A+1.1}{A+4}$ &$\frac{0.36 A+0.03}{A-0.02}$\\
    &$3 s    $  & $1$ & $\frac{0.04A-0.14}{A-1.6}$ & $\frac{0.12 A-0.09}{A-1.6}$& $\frac{0.56 A-0.91}{A-2.9}$& $\frac{-27 A+92}{160}$ &$\frac{  0.3 A-0.24}{A+0.54}$\\
    &$3\sigma     $  & $\frac{-0.16 A+0.16}{A-1.3}$ & $1$ & $\frac{0.17 A-0.17}{A-1.3}$& $\frac{0.11A-0.11}{A-1}$& $\frac{0.39A-0.49}{A-1.3}$ &$\frac{0.33A-0.33}{A-1}$\\
    &$4      $  & $1$ & $\frac{0.03 A+0.11}{A+0.16}$ & $\frac{-0.1A-0.03}{A+0.61}$& $\frac{0.63A+0.38}{A+1.3}$& $\frac{-0.82A-3.1}{A-3.9}$ &$\frac{1.2A+0.34}{A+4.1}$\\\hline\hline
  \end{tabular}
  \end{center}
  \end{table}
\begin{table}\centering
  \caption{Input parameters of the singly and doubly heavy baryons and heavy mesons}\label{tab:inputparameter}
\begin{tabular}{c|c|c|c|c|c}
    \hline     \hline
    hadron &  mass & decay constant & hadron & mass & decay constant \\\hline 
    $D$ & 1.86 & 0.1904 & $B$ & 5.28 & 0.2011\\ 
    $D_s$ & 1.97 & 0.2007 & $B_s$ & 5.37 & 0.2232\\
    $D^{*}$ & 2.01 & 0.2455 & $B^{*}$ & 5.32 & 0.2142\\
    $D_s^{*}$ & 2.11 & 0.2811 & $B_s^{*}$ & 5.42 & 0.2452\\
    \hline
    $\Xi_{cc}$ & 3.621 & 0.109 & $\Omega_{cc}$ & 3.690 & 0.123 \\ 
    $\Xi_{bc}$& 6.89 & 0.176 & $\Omega_{bc}$& 6.95 & 0.188  \\ 
    $\Xi_{bb}$ & 10.143 & 0.281 &$\Omega_{bb}$ & 10.19 & 0.347 \\\hline 
    ${\Lambda}_{c}$ & 2.29 & 0.022&${\Lambda}_{b}$ & 5.62 & 0.030\\
    ${\Xi}_{c}$ & 2.47 & 0.027&${\Xi}_{b}$ & 5.80 & 0.032\\
    \hline \hline
    \end{tabular}
  \end{table}

  In conclusion, the sum rules for coupling constants incorporate two auxiliary parameters: the Borel parameter $M^{2}_{1,2}$ and the continuum threshold $s_{\rm th}$. It is imperative to identify the operational regions of these parameters where variations in the results of coupling constants are minimal despite changes in these parameters. To constrain these parameters, we adhere to the standard methodologies, including pole dominance, convergence of the OPE, and moderate fluctuations of physical quantities relative to the auxiliary parameters.
  The Borel parameters $M^{2}_{1,2}$ and threshold $s_{\rm th}$ for ${\cal B}_{QQ^{\prime}}$ are chosen carefully, which are listed in Tab.~\ref{Tab:para_Mands} and the uncertainties of them are taken approximately $10\%$ of their central values. Given these input parameters, we can calculate the numerical results for the strong coupling constants $g_{P{\cal B}_{Q}{\cal B}_{QQ^{\prime}}}$ and $f_{V{\cal B}_{Q}{\cal B}_{QQ^{\prime}}}$ as defined in Eqs.~(\ref{eq:hqgpbbF})-(\ref{eq:hqf2vbbF}).
\begin{table}[!htb]
\caption{The Borel parameter and continuum threshold of doubly heavy baryons
for strong coupling constant.}
\label{Tab:para_Mands} 
\begin{tabular}{c|c|c|c}
\hline \hline
{ Coupling} & {$s_{\rm th}$ (GeV$^{2}$) } & {$M^{2}_{2}$ (GeV$^{2}$) }& {$M^{2}_{1}$ (GeV$^{2}$) } \tabularnewline
\hline 
{{${\cal B}_{cc}{\cal B}_{c}D$}} & {{$15\sim 19$}} & {{$1.6\sim 2.0$}}& {{$14\sim 16$}}\tabularnewline\hline
{{${\cal B}_{bc}{\cal B}_{b}D$}} & {$51\sim57$ } & {$5\sim 6$ } & {{$45\sim55$}}\tabularnewline\hline
{{${\cal B}_{bc}{\cal B}_{c}B$}} & {$51\sim57$ } & {$2.5\sim 3.5$ } & {{$45\sim55$}}\tabularnewline
\hline 
{{${\cal B}_{bb}{\cal B}_{b}B$}} & {{$107\sim117$}} & {{$8.5\sim 10.5$}} & {{$90\sim110$}}\tabularnewline\hline\hline
\end{tabular}
\end{table}

\subsection{Strong coupling constants}
\begin{figure}[t]
  \centering
  \includegraphics[width=0.45\textwidth]{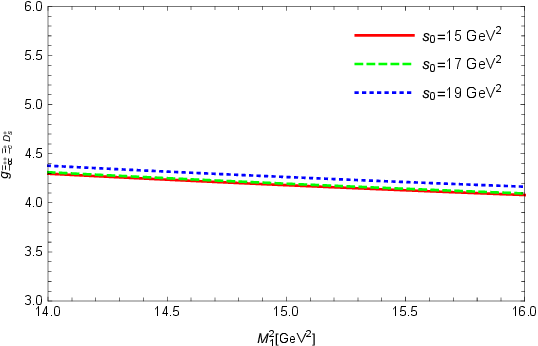}
  \includegraphics[width=0.45\textwidth]{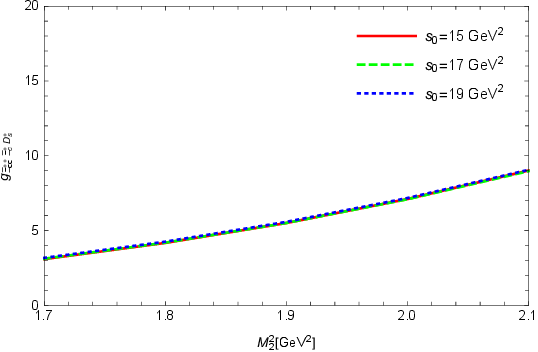}\\
  \includegraphics[width=0.45\textwidth]{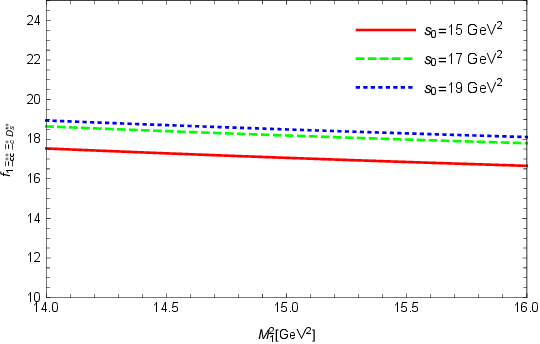}
  \includegraphics[width=0.45\textwidth]{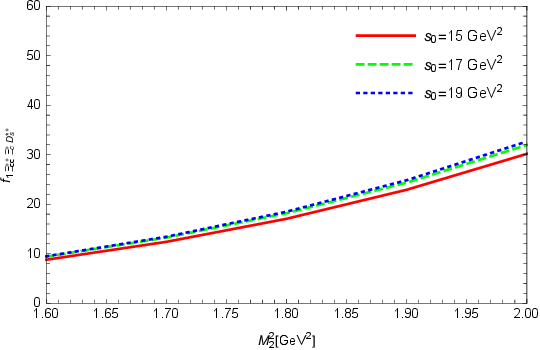}\\
  \includegraphics[width=0.45\textwidth]{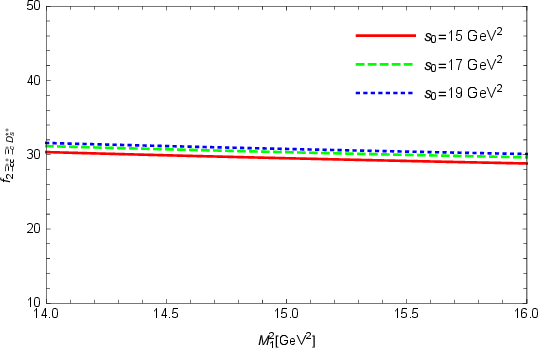}
  \includegraphics[width=0.45\textwidth]{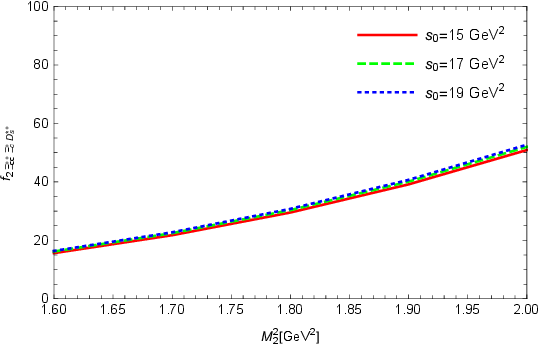}\\
  \includegraphics[width=0.45\textwidth]{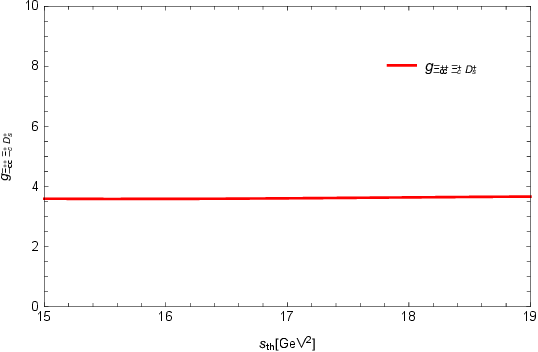}
  \includegraphics[width=0.45\textwidth]{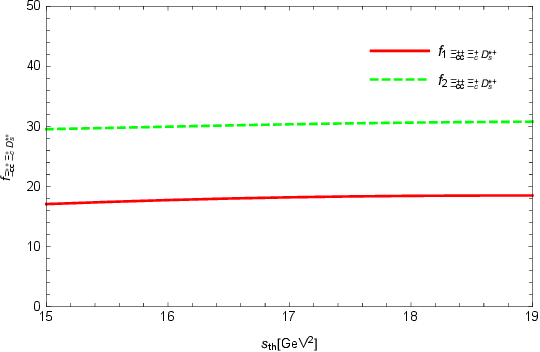}\\
     \caption{The strong couplings $g_{\Xi^{++}_{cc}\Xi^{+}_{c} D}$ and $f_{\Xi^{++}_{cc}\Xi^{+}_{c} D^{*}}$ as the functions of the Borel
parameters $M^2_{1}$ and $M^2_{2}$ for different values of $s_{\rm th}$.}
\label{fig:fig2}
\end{figure}

From the Fig.~\ref{fig:fig2}, we observe mild variations in 
$g_{\Xi^{++}_{cc}\Xi^{+}_{c} D}$ and $f_{\Xi^{++}_{cc}\Xi^{+}_{c} D^{*}}$ with respect to the Borel parameter $M^2_{1,2}$ and continuum threshold 
$s_{\rm th}$, which constitute the primary source of uncertainty in the numerical values of the strong coupling constants. The extracted numerical values for the strong couplings $g_{\Xi^{++}_{cc}\Xi^{+}_{c} D}$ and $f_{\Xi^{++}_{cc}\Xi^{+}_{c} D^{*}}$ are presented in Tab.~\ref{tab:cccD}. The reported errors arise from variations in the Borel parameter $M^2_{1,2}$ within their working regions, as well as from uncertainties of the continuum threshold 
$s_{\rm th}$.
From Fig.~\ref{fig:fig2} and Tab.~\ref{tab:cccD}, we can found that
\begin{itemize}
  \item The errors of the strong couplings given in Tab.~\ref{tab:cccD} mainly results from the errors of the Borel parameter $M^2_{2}$.
  This is because the integration limit of $s^{\prime}$ is dependence on the integration of $s$. And after double integration of $s$ and $s^{\prime}$, the reslut is dependence on the $e^{M^2_{2}/M_{M}^2}$.
  \item According to the SU(3) symmetry, the various strong couplings are
  related with each other. In this work, we have offered a systematic SU(3) analysis of the strong couplings ${\cal B}_{cc}{\cal B}_{c}D^{(*)}$, ${\cal B}_{bc}{\cal B}_{b}D^{(*)}$, ${\cal B}_{bc}{\cal B}_{c}B^{(*)}$ and ${\cal B}_{bb}{\cal B}_{b}B^{(*)}$. And the LCSR results are the same with the relations derived by SU(3) flavor symmetry. 
  \item As shown in Tab.~\ref{tab:cccD}, the strong couplings of ${\cal B}_{QQ^{\prime}}{\cal B}_{Q}V_{Q^{\prime}}$ is much large than the ones of ${\cal B}_{QQ^{\prime}}{\cal B}_{Q}P_{Q^{\prime}}$.
\end{itemize}

\begin{table}
  \caption{The numerical results of strong couplings ${\cal B}_{QQ^{\prime}}{\cal B}_{Q}M_{Q^{\prime}}$ can be calculated using the Eqs.~(\ref{eq:hqgpbbF}-\ref{eq:hqf2vbbF}). And the uncertainty associated with the Borel parameters \(M^{2}_{1,2}\) and the threshold \(s_{\rm th}\) is estimated to be approximately 10\% of their central values.}\label{tab:cccD}
  \begin{tabular}{l|c|c|c|c|c|c}
    \hline\hline
    Coupling &SU(3)&$g$&Coupling &SU(3)&$f_{1}$&$f_{2}$
    \\ \hline
    $\Xi_{cc}^{++}\Lambda_{c}^{+}D^{+}$ &$a_{1}$&$2.72\pm0.92$&
    $\Xi_{cc}^{++}\Lambda_{c}^{+}D^{*+}$ &$b_{1}$&$12.20\pm4.60$&
    $20.00\pm7.06$
    \\
    $\Xi_{cc}^{++}\Xi_{c}^{+}D_{s}^{+}$ &$a_{1}$&$4.19\pm1.39$&
    $\Xi_{cc}^{++}\Xi_{c}^{+}D_{s}^{*+}$ &$b_{1}$&$18.20\pm6.67$&
    $30.30\pm10.40$
    \\
    $\Xi_{cc}^{+}\Lambda_{c}^{+}D^{0}$ &$-a_{1}$&$-2.70\pm0.92$&
    $\Xi_{cc}^{+}\Lambda_{c}^{+}D^{*0}$ &$-b_{1}$&$-12.20\pm4.58$&
    $-19.80\pm7.03$
    \\
    $\Xi_{cc}^{+}\Xi_{c}^{0}D_{s}^{+}$ &$a_{1}$&$4.20\pm1.39$&
    $\Xi_{cc}^{+}\to\Xi_{c}^{0}D_{s}^{*+}$ &$b_{1}$&$18.20\pm6.66$&
    $30.30\pm10.40$
    \\
    $\Omega_{cc}^{+}\Xi_{c}^{+}D^{0}$ &$-a_{1}$&$-3.62\pm1.26$&
    $\Omega_{cc}^{+}\to\Xi_{c}^{+}D^{*0}$ &$-b_{1}$&$-15.80\pm6.11
     
      $&
    $-26.60\pm9.64$
    \\
    $\Omega_{cc}^{+}\Xi_{c}^{0}D^{+}$ &$-a_{1}$&$-3.64\pm1.27$&
    $\Omega_{cc}^{+}\to\Xi_{c}^{0}D^{*+}$ &$-b_{1}$&$-15.90\pm6.13
     
      $&
    $-26.80\pm9.68$
    \\\hline
    $\Xi_{bc}^{+}\Lambda_{c}^{+}\bar B^{0}$ &$a_{2}$&$1.19\pm0.74$&
    $\Xi_{bc}^{+}\to\Lambda_{c}^{+}\bar B^{*0}$ &$b_{2}$&$13.50\pm
     
      9.54$&
    $16.80\pm11.50$
    \\
    $\Xi_{bc}^{+}\Xi_{c}^{+}\bar B_{s}^{0}$ &$a_{2}$&$2.21\pm1.29$&
    $\Xi_{bc}^{+}\to\Xi_{c}^{+}\bar B_{s}^{*0}$ &$b_{2}$&$23.60\pm
     
      15.60$&
    $29.30\pm18.80$
    \\
    $\Xi_{bc}^{0}\Lambda_{c}^{+}B^{-}$ &$-a_{2}$&$-1.19\pm0.74$&
    $\Xi_{bc}^{0}\to\Lambda_{c}^{+}B^{*-}$ &$-b_{2}$&$-13.50\pm9.54
     
      $&
    $-16.80\pm11.50$
    \\
    $\Xi_{bc}^{0}\Xi_{c}^{+}\bar B_{s}^{0}$ &$a_{2}$&$2.21\pm1.30$&
    $\Xi_{bc}^{0}\to\Xi_{c}^{+}\bar B_{s}^{*0}$ &$b_{2}$&$23.60\pm
     
      15.60$&
    $29.30\pm18.80$
    \\
    $\Omega_{bc}^{0}\Xi_{c}^{+}B^{-}$ &$-a_{2}$&$-1.77\pm1.12$&
    $\Omega_{bc}^{0}\to\Xi_{c}^{+}B^{*-}$ &$-b_{2}$&$-18.90\pm13.50
     
      $&
    $-23.60\pm16.40$
    \\
    $\Omega_{bc}^{0}\Xi_{c}^{0}\bar B^{0}$ &$-a_{2}$&$-1.77\pm1.12
     
      $&
    $\Omega_{bc}^{0}\to\Xi_{c}^{0}\bar B^{*0}$ &$-b_{2}$&$-18.90\pm
     
      13.50$&
    $-23.60\pm16.40$
          \\ \hline
          $\Xi_{bc}^{+}\Lambda_{b}^{0}D^{+}$ &$a_{3}$&$1.51\pm0.62$&
          $\Xi_{bc}^{+}\to\Lambda_{b}^{0}D^{*+}$ &$b_{3}$&$6.75\pm2.84$&
          $15.20\pm6.32$
          \\
          $\Xi_{bc}^{+}\Xi_{b}^{0}D_{s}^{+}$ &$a_{3}$&$1.69\pm0.69$&
          $\Xi_{bc}^{+}\to\Xi_{b}^{0}D_{s}^{*+}$ &$b_{3}$&$7.40\pm3.11$&
          $16.70\pm6.93$
          \\
          $\Xi_{bc}^{0}\Lambda_{b}^{0}D^{0}$ &$-a_{3}$&$-1.51\pm0.62$&
          $\Xi_{bc}^{0}\to\Lambda_{b}^{0}D^{*0}$ &$-b_{3}$&$-6.75\pm2.84
           
            $&
          $-15.20\pm6.31$
          \\
          $\Xi_{bc}^{0}\Xi_{b}^{-}D_{s}^{+}$ &$a_{3}$&$1.69\pm0.69$&
          $\Xi_{bc}^{0}\to\Xi_{b}^{-}D_{s}^{*+}$ &$b_{3}$&$7.39\pm3.11$&
          $16.70\pm6.93$
          \\
          $\Omega_{bc}^{0}\Xi_{b}^{0}D^{0}$ &$-a_{3}$&$-1.76\pm0.72$&
          $\Omega_{bc}^{0}\to\Xi_{b}^{0}D^{*0}$ &$-b_{3}$&$-7.84\pm3.30$&
          $-17.80\pm7.40$
          \\
          $\Omega_{bc}^{0}\Xi_{b}^{-}D^{+}$ &$-a_{3}$&$-1.75\pm0.72$&
          $\Omega_{bc}^{0}\to\Xi_{b}^{-}D^{*+}$ &$-b_{3}$&$-7.84\pm3.30$&
          $-17.80\pm7.40$
            \\ \hline
            $\Xi_{bb}^{0}\Lambda_{b}^{0}\bar B^{0}$ &$a_{4}$&$1.58\pm0.50$&
            $\Xi_{bb}^{0}\to\Lambda_{b}^{0}\bar B^{*0}$ &$b_{4}$&$15.60\pm
             
              5.07$&
            $24.80\pm7.96$
            \\
            $\Xi_{bb}^{0}\Xi_{b}^{0}\bar B_{s}^{0}$ &$a_{4}$&$1.92\pm0.61$&
            $\Xi_{bb}^{0}\to\Xi_{b}^{0}\bar B_{s}^{*0}$ &$b_{4}$&$18.60\pm
             
              6.04$&
            $29.70\pm9.52$
            \\
            $\Xi_{bb}^{-}\Lambda_{b}^{0}B^{-}$ &$-a_{4}$&$-1.58\pm0.50$&
            $\Xi_{bb}^{-}\to\Lambda_{b}^{0}B^{*-}$ &$-b_{4}$&$-15.60\pm5.07
             
              $&
            $-24.80\pm7.96$
            \\
            $\Xi_{bb}^{-}\Xi_{b}^{-}\bar B_{s}^{0}$ &$a_{4}$&$1.92\pm0.61$&
            $\Xi_{bb}^{-}\to\Xi_{b}^{-}\bar B_{s}^{*0}$ &$b_{4}$&$18.60\pm
             
              6.04$&
            $29.70\pm9.52$
            \\
            $\Omega_{bb}^{-}\Xi_{b}^{0}B^{-}$ &$-a_{4}$&$-1.73\pm0.55$&
            $\Omega_{bb}^{-}\to\Xi_{b}^{0}B^{*-}$ &$-b_{4}$&$-16.90\pm5.52
             
              $&
            $-27.60\pm8.89$
            \\
            $\Omega_{bb}^{-}\Xi_{b}^{-}\bar B^{0}$ &$-a_{4}$&$-1.73\pm0.55
             
              $&
            $\Omega_{bb}^{-}\to\Xi_{b}^{-}\bar B^{*0}$ &$-b_{4}$&$-16.90\pm
             
              5.52$&
            $-27.60\pm8.89$
            \\\hline\hline
            \end{tabular}   \end{table}
\section{Conclusions}
\label{sec:conclusions}

This study presents the first LCSRs calculation of the strong coupling constants ${\cal B}_{QQ^{\prime}}{\cal B}_{Q}P_{Q^{\prime}}$ and ${\cal B}_{QQ^{\prime}}{\cal B}_{Q}V_{Q^{\prime}}$, where $Q,Q^{\prime}$ represent either $c$ or $b$ quark. We employ the general form of interpolating currents for doubly heavy baryons and heavy mesons, along with the light-cone distribution amplitudes (DAs) of singly heavy baryons anti-triplet. Following the standard prescriptions of the method, we fixed the auxiliary parameters involved in the calculations and extracted the values of the different strong coupling constants. Our results can be verified using other theoretical models and approaches, like QCD Sum Rules and may contribute to constructing the strong interaction potential between doubly heavy baryons and heavy mesons. Additionally, these findings could assist experimental groups in analyzing related data from hadron colliders.

Quark-hadron duality was employed to derive the strong couplings, and the numerical results are provided in Tab.~\ref{tab:cccD}. These results can serve as inputs for future research on non-leptonic decays. Detailed error estimations and theoretical analyses are also provided. Furthermore, Tab.~\ref{tab:cccD} includes a set of SU(3) relations between different couplings for comparison. We hope that this investigation will provide valuable inputs for theoretical and experimental studies of doubly heavy baryons conducted by the LHCb and other experiments, contributing to a deeper understanding of the dynamics of baryon decays.

\section*{Acknowledgements}

The authors are very grateful to Prof.~Fu Sheng Yu and Dr. Hua Yu Jiang for useful discussions.
This work is supported in part by the youth Foundation JN210003, of
China University of mining and technology, the 13th Sailing Plan No.102521101,
of China University of mining and technology, the opening Issues Foundation
No.12247101, of Lanzhou Center for Theoretical Physics, and the National
Natural Science Foundation of China under Grants No. 12005294 and No.12305103.

\end{document}